\documentclass[a4paper,11pt]{article}
\pdfoutput=1 
\usepackage{jinstpub}
\usepackage{siunitx}

\title{\boldmath Improving sensitivity of the ARIANNA detector by rejecting thermal noise with deep learning}

\author[a,1]{A.~Anker \note{Corresponding author}}
\author[b]{P. Baldi}
\author[a]{S. W. Barwick}
\author[c]{J. Beise}
\author[d]{D. Z. Besson}
\author[e]{S. Bouma}
\author[e]{M. Cataldo}
\author[f]{P. Chen}
\author[a]{G. Gaswint}
\author[c]{C. Glaser}
\author[c]{A. Hallgren}
\author[g]{S. Hallmann}
\author[h]{J. C. Hanson}
\author[i]{S. R. Klein}
\author[j]{S. A. Kleinfelder}
\author[e]{R. Lahmann}
\author[a]{J. Liu}
\author[d]{M. Magnuson}
\author[b]{S. McAleer}
\author[g, e]{Z. S. Meyers}
\author[f]{J. Nam}
\author[e,g]{A. Nelles}
\author[d]{A. Novikov}
\author[a]{M. P. Paul}
\author[a]{C. Persichilli}
\author[e,g]{I. Plaisier}
\author[g, e]{L. Pyras}
\author[a]{R. Rice-Smith}
\author[k]{J. Tatar}
\author[f]{S.-H Wang}
\author[e,g]{C. Welling}
\author[a]{L. Zhao}

\affiliation[a]{Department of Physics and Astronomy, University of California, Irvine, CA 92697, USA.}
\affiliation[b]{Department of Information and Computer Science, University of California, Irvine, CA 92697, USA.}
\affiliation[c]{Uppsala University Department of Physics and Astronomy, Uppsala SE-752 37, Sweden.}
\affiliation[d]{Department of Physics and Astronomy, University of Kansas, Lawrence, KS 66045, USA.}
\affiliation[e]{ECAP, Friedrich-Alexander Universität Erlangen-Nürnberg, 91058 Erlangen, Germany.}
\affiliation[f]{Department of Physics and Leung Center for Cosmology and Particle Astrophysics, National Taiwan University, Taipei 10617, Taiwan.}
\affiliation[g]{DESY, 15738 Zeuthen, Germany. }
\affiliation[h]{Whittier College Department of Physics, Whittier, CA 90602, USA.}
\affiliation[i]{Lawrence Berkeley National Laboratory, Berkeley, CA 94720, USA.}
\affiliation[j]{Department of Electrical Engineering and Computer Science, University of California, Irvine, CA 92697, USA.}
\affiliation[k]{Research Cyberinfrastructure Center, University of California, Irvine, CA 92697, USA. }

\emailAdd{aanker@uci.edu, mppaul@uci.edu, sbarwick@uci.edu, christian.glaser@physics.uu.se}

\abstract{The ARIANNA experiment is an Askaryan detector designed to record radio signals induced by neutrino interactions in the Antarctic ice. Because of the low neutrino flux at high energies ($E_{\nu}> \SI{e16}{eV}$), the physics output is limited by statistics. Hence, an increase in sensitivity significantly improves the interpretation of data and offers the ability to probe new parameter spaces. The amplitudes of the trigger threshold are limited by the rate of triggering on unavoidable thermal noise fluctuations. We present a real-time thermal noise rejection algorithm that enables the trigger thresholds to be lowered, which increases the sensitivity to neutrinos by up to a factor of two (depending on energy) compared to the current ARIANNA capabilities. A deep learning discriminator, based on a Convolutional Neural Network (CNN), is implemented to identify and remove thermal events in real time. We describe a CNN trained on MC data that runs on the current ARIANNA microcomputer and retains 95\% of the neutrino signal at a thermal noise rejection factor of $10^5$, compared to a template matching procedure which reaches only $10^2$ for the same signal efficiency. Then the results are verified in a lab measurement by feeding in generated neutrino-like signal pulses and thermal noise directly into the ARIANNA data acquisition system. Lastly, the same CNN is used to classify cosmic-rays events to make sure they are not rejected. The network classified 102 out of 104 cosmic-ray events as signal. 
}

\collaboration{ARIANNA collaboration}

\begin{document}
\maketitle
\flushbottom

\section{Introduction}
\label{sec:intro}

Ultra-high-energy (UHE, defined here as $E_{\nu}> \SI{e17}{eV}$) neutrino astronomy expands the opportunity to learn more about the fierce processes of astronomical objects \cite{Ackermann:2019ows}. Neutrinos are ideal messengers because they have negligible mass, are neutral in charge, and, due to the fact that they only interact through the weak force,  have a low interaction probability.  Once created,  these properties allow them to travel through space unhindered by intervening matter or radiation such as dust, gas, and electromagnetic fields. The same properties also make them challenging to detect. Even at  the extreme energies relevant to radio neutrino detectors, neutrinos rarely interact with matter. When this feature is combined with the low expected fluxes, and stringent experimental upper limits have been published by the IceCube Collaboration \cite{Aartsen_2018}, the detector architecture must incorporate  large volumes of target material. A rough estimate suggests that instrumented volumes must reach of order one teraton ($10^{12}$ $m^{3}$) to observe a few neutrinos per year for commonly discussed theoretical models of neutrino production \cite{van_Vliet_2019}. 

Radio based neutrinos experiments have been successfully explored in the past with pilot arrays such as the ARA experiment \cite{ARA} and the ARIANNA experiment \cite{ARIANNA}, the latter being the focal point of this paper. These efforts helped focus in on the radio techniques required to operate in extremely cold and harsh conditions. While these experiments showed the technical feasibility, they were too small to measure the low neutrino flux. Undeterred, several radio-based experiments in development are further illustrating the capabilities of this detection method, such as ARIANNA-200 \cite{anker2020white}, the radio component of IceCube-Gen2 \cite{Gen2WhitePaper}, the Radio Neutrino Observatory in Greenland (RNO-G) \cite{RNOG}, Giant Radio Array for Neutrino Detection (GRAND) \cite{GRAND}, Taiwan Astroparticle Radiowave Observatory for Geo-synchrotron Emissions (TAROGE) \cite{TAROGE}, and Payload for Ultrahigh Energy Observations (PUEO) \cite{PUEO}, a successor to ANITA \cite{ANITA}. These experiments exploit various target materials such as ice, water, mountains, and air.

The challenge for experimenters is to reach the teraton detection volumes at a reasonable cost. One of the most promising methods for observing UHE neutrinos in large target volumes exploits radio detection in ice \cite{Schroder:2016hrv,Price:1995ep}. For this reason, locations such as Greenland and Antarctica are popular sites for radio detection experiments. Ice is transparent to radio signals, with field attenuation lengths ranging from \SI{0.5}{km} at Moore's Bay (Antarctica) \cite{BarwickBergBessonEtAl2014} to more than a kilometer in colder ice found at the South Pole \cite{barwick_besson_gorham_saltzberg_2005} or the Greenland ice sheet \cite{Avva:2014ena}. Radio pulses are created via the Askaryan effect \cite{Askaryan:1962hbi} when interacting neutrinos create particles showers in ice, which in turn generate a time-varying negative charge excess that produces radio emission in the \SI{50}{MHz} to \SI{1}{GHz} range. 

The radio technique enables cost-efficient instrumentation for monitoring large detection volumes. However, because of the low flux of UHE neutrinos, event rates are still small even for the large array of hundreds of radio detector stations that is foreseen for the next-generation neutrino observatory at the South Pole, IceCube-Gen2 \cite{HallmannICRC2021}. Thus, improving the sensitivity of the detector is one of the primary objectives. The easiest way to increase the sensitivity -- but also the most expensive way -- is to build more radio detector stations. A more efficient way is to increase the sensitivity of each radio detector station and a lot of work has been made towards this goal. 

The sensitivity can be increased by simply lowering the trigger threshold which records additional neutrino interactions that produce smaller signal strengths in the radio detector. The problem with this is that the trigger thresholds are already set close to the thermal noise floor such that the trigger rate is dominated by unavoidable thermal noise fluctuations. The trigger rate on thermal noise fluctuations changes drastically with threshold. For example, an amplitude threshold trigger with a two out of four antenna coincidence logic has a trigger rate increases by about six orders-of-magnitude if the trigger threshold is lowered from four times the RMS noise, $V_{\text{RMS}}^{\text{noise}}$, to just three time $V_{\text{RMS}}^{\text{noise}}$ \cite{Glaser2020Bandwidth}. Therefore, the trigger threshold is limited by the maximum data rate a radio detector can handle which is typically on the order of \SI{1}{Hz} if a high-speed communication link exists. If the communication relies on Iridium satellite communication, the maximum data rate is limited to \SI{0.3}{mHz}. However, if thermal noise fluctuations are identified and rejected in real time, the trigger thresholds can be lowered while maintaining the same data rate, thus increasing the sensitivity of the detector. The sensitivity can be improved by up to a factor of two with the intelligent trigger system presented here (cf. Sec.~\ref{subsection:sensitivity}).

In this paper it is demonstrated that deep learning can be used to reject thermal noise in real time by implementing these techniques in the current ARIANNA data acquisition system. Deep learning, a modern rebranding of neural networks, has been shown to outperform other methods in a variety of scientific and engineering areas, including in physics \cite{dl-baldi2014,baldi2021deep}. The significant amount of data that need to be classified in real time with low latency in high energy physics experiments makes deep learning an ideal tool to use \cite{Duarte_2018,kronmueller2019application}. By rejecting thermal events, the trigger rate can be increased dramatically while maintaining the required low rate of event transmission over the communication links from the remotely located ARIANNA stations. Overall, lower thresholds increase the effective volume of ice observed by each station, which is proportional to the sensitivity of the detector. 

This paper is organized as follows. Additional details on the ARIANNA detector are provided, along with the expected gain in sensitivity for this study. Next the trade off between network efficiency and processing time is assessed to find the optimal deep learning models for a representative sample of microprocessor platforms. The deep learning method is then compared to a template matching study to determine how well the more common approach performs. Then the current ARIANNA data acquisition system is evaluated to determine the suitability for a deep learning filter. Moreover, the specific predictions for the optimal deep learning model are experimentally verified for the current microprocessor hardware. Lastly, the deep learning filter is tested on measured cosmic rays to verify that they are classified similar to neutrino signal and not rejected as thermal noise. The paper concludes with a short summary and plans for the future.

\section{The ARIANNA experiment and expected gain in sensitivity } 
The ARIANNA experiment \cite{barwick2014design} is an array of autonomous radio stations located in Antarctica. Stations have operated at sea-level on the Ross Ice Shelf in Moore's Bay, about \SI{110}{km} from McMurdo Station, which is the largest research base on the continent. In addition, two stations have operated at the South Pole, which is colder and higher in elevation than the environment at Moore's Bay.

\subsection{Detector description}
\begin{figure}
{\centering
\includegraphics[width=0.55\textwidth]{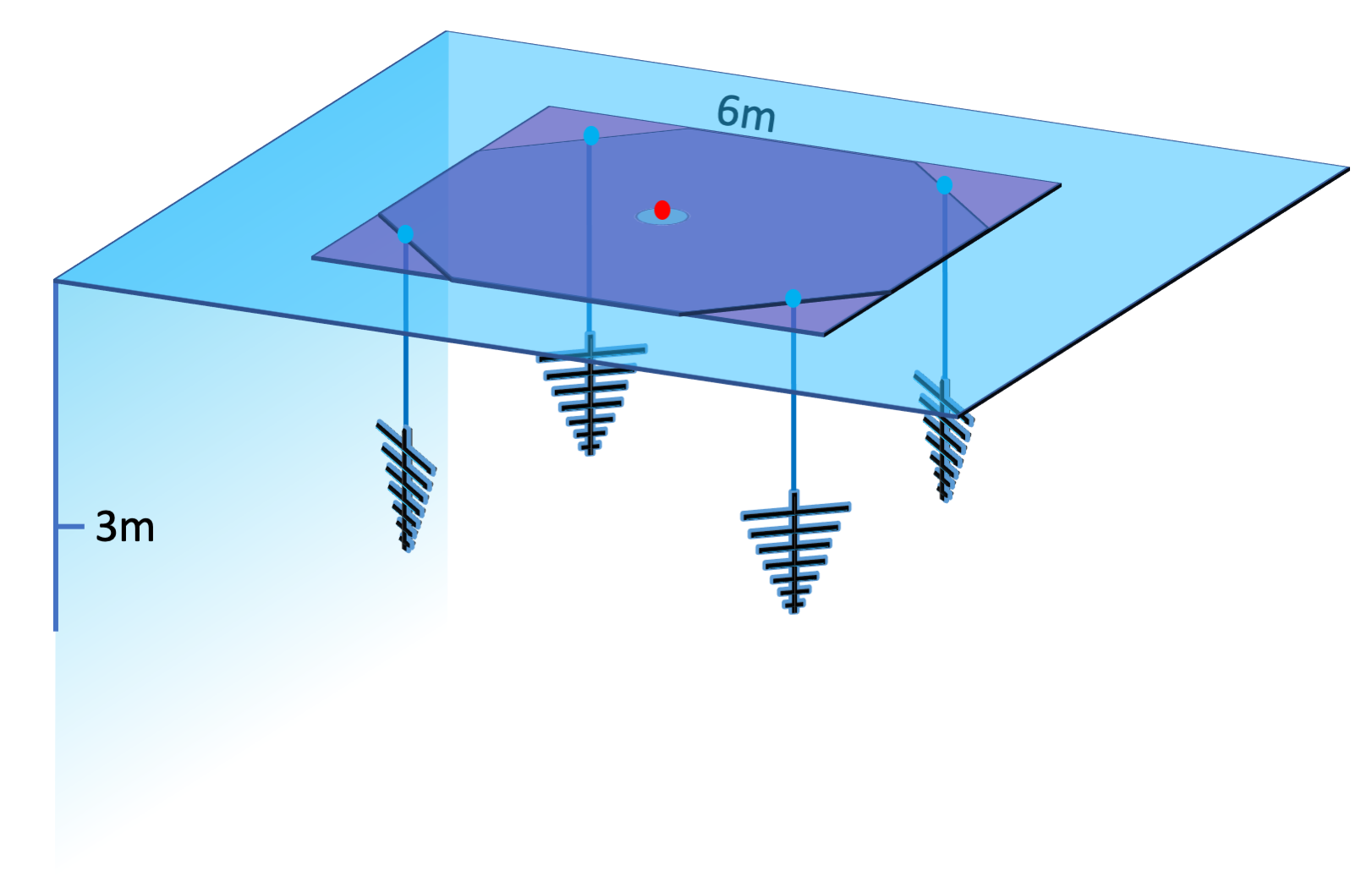}
\caption{Diagram of a typical ARIANNA station consisting of four downward facing log periodic dipole antennas (LPDAs) located three meters below the Antarctic Ice.}
\label{fig:stn_dgrm}
}
\end{figure}

Several architectures were implemented in the prototype array at Moore's Bay. Most stations consisted of four downward facing log periodic dipole antennas (LPDAs) to specifically look for neutrino events, as shown in Fig.~\ref{fig:stn_dgrm}. Two other stations at Moore's Bay and two at the South Pole were configured with eight antennas, which included a mixture of LPDAs and dipoles. These stations were simultaneously sensitive to cosmic rays that interact in the atmosphere \cite{Barwick2017-Airshowers,leshanICRC} and neutrinos. The radio signals are digitized and captured using a custom-made chip design known as the SST \cite{Kleinfelder_2014}. The analog trigger system of ARIANNA imposes requirements on individual waveforms; a high and low threshold must occur within \SI{5}{ns}, and multiple antennas channels (at least two of four antennas) must meet the high-low threshold within a \SI{30}{ns} coincidence window. These criteria are based on the expectation that thermal noise fluctuations are approximately independent, whereas neutrino signals produce correlated high-low fluctuations in a given antenna, and produce comparable signals in multiple antenna channels. These requirements reduce the rate of thermal noise triggers for a given trigger threshold while maintaining the sensitivity to Askaryan pulses from high-energy neutrinos. Once a station has triggered, the digitized waveforms of every antenna channel contain 256 samples with a voltage accuracy of \SI{12}{bits}. The event size in an eight-channel station is \SI{132}{kbits}. The waveform data from all channels are piped into an Xilinx Spartan 4 FPGA, and then further processed and stored to an internal \SI{32}{GB} memory card by an MBED LPC 1768 microcontroller. There are up to eight channels on each board that process the radio signal from each antenna. 

Once a triggered event is saved to local storage (the memory card) it is then transferred to UC Irvine through a long-range WiFi link (AFAR) \cite{AFAR} during a specified communication window. The ARIANNA stations also use Iridium satellite network as a backup system. Satellite communication is relatively slow, with a typical transfer rate of one event every \SI{2}-\SI{3}{minutes} \cite{ggaswint_thesis}. For both communication methods, currently the hardware system is limited to either communication or data collection. Therefore, neutrino search operations are disabled during data communication. As radio neutrino technologies move beyond the prototype stage, the relatively expensive and power consumptive AFAR system will be eliminated. Perhaps it will be replaced by a better wireless system, such as LTE, for sites relatively close to scientific research bases, but for more remote locations, only satellite communications such as Iridium are feasible. Given the current limitation of \SI{0.3}{events/min} imposed by Iridium communication, and the fact that neutrino operations cease during data transfer which generates unwanted deadtime, stations that rely solely on Iridium communication are expected to operate at trigger rates from $\sim$\SI{0.3}{mHz} to keep losses due to data transfer, $f_{\text{trans}}$, below 3\%.  

The trigger thresholds of ARIANNA are adjusted to a certain multiple of the Signal to Noise Ratio (SNR), defined here as the ratio of the maximum absolute value of the amplitude of the waveform to the $V_{\text{RMS}}^{\text{noise}}$. Currently, the pilot stations are set to trigger above 4.4 SNR to reach the constrained trigger rate of order \SI{1}{mHz}. In the next section, the expected gain in sensitivity is studied for a lower threshold of 3.6 SNR, which corresponds to \si{100}{Hz}, the maximum operation rate of the stations. For more information on the ARIANNA detector, see \cite{anker2020white,Anker_2019}.

\subsection{Expected gain in sensitivity} 
\label{subsection:sensitivity}
The real-time rejection of thermal noise that is presented in this article would enable the trigger threshold to be lowered significantly -- thus increasing the detection rate of UHE neutrinos -- while keeping a low event rate of a few \si{mHz}. To estimate the increase in sensitivity, the effective volume of an ARIANNA station is simulated for the two trigger thresholds corresponding to a thermal noise trigger rate of \SI{10}{mHz} (the current ARIANNA capabilities), and a four orders-of-magnitude higher trigger rate (enabled through the deep-learning filter that rejects 99.99\% of all thermal noise triggers). We use the relationship between trigger threshold and trigger rate from \cite{Glaser2020Bandwidth} to calculate the thresholds. 

\begin{figure}[t]
{\centering
\includegraphics[width=0.55\textwidth]{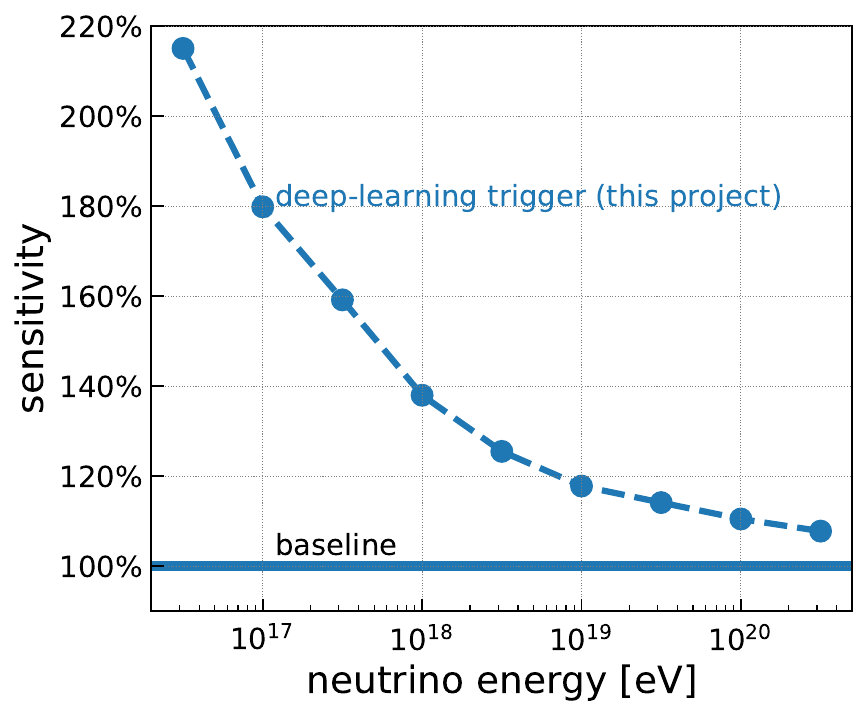}
\caption{Expected improvement in sensitivity to high-energy neutrinos with the deep-learning trigger developed in this work. The improvement in sensitivity directly translates into the number of observed neutrinos. The baseline is the standard ARIANNA high/low trigger with a 2 out of 4 antennas coincidence requirement for the nominal bandwidth of \SIrange{80}{800}{MHz} at a thermal noise trigger rate of \SI{10}{mHz}. The blue dashed curve shows the sensitivity for a trigger threshold corresponding to a trigger rate of \SI{100}{Hz} and otherwise the same simulation settings.}
\label{fig:improv_sens}
}
\end{figure}

NuRadioMC \cite{NuRadioMC} is used to simulate the sensitivity of the ARIANNA detector at Moore's Bay. The expected radio signals are simulated in the ARIANNA detector on the Ross ice shelf, i.e., an ice shelf with a thickness of \SI{576}{m} and an average attenuation length of approx. \SI{500}{m}, and where the ice-water interface at the bottom of the ice shelf reflects radio signals back up with high efficiency. The generated neutrino interactions are distributed uniformly in the ice around the detector with random incoming directions. The simulation is performed for discrete neutrino energies and includes a simulation of the full detector response and the trigger algorithm as described above. The resulting gain in sensitivity is shown in Fig.~\ref{fig:improv_sens} and increases by almost a factor of two at energies of \SI{e17}{eV}. The improvement decreases towards higher energies because fewer of the recorded events are close to the trigger threshold but at \SI{e18}{eV} there is still an increase in sensitivity of 40\%.

\section{Thermal noise rejection using deep neural networks}
\label{sec:dl_arch}
To implement a deep learning filter, the general network structure needs to be optimized for fast and accurate classification. For accuracy, the two metrics are neutrino signal efficiency (defined here as the ratio of correctly identified signal events to the total number of signal events) and noise rejection factor (defined here as $\frac{1}{(1-N_{\text{ratio}})}$, where $N_{\text{ratio}}$ is the ratio of correctly identified noise events to the total number of noise events). The goal is to reject several orders-of-magnitude of thermal noise fluctuations while retaining most of the neutrino signals. In the following, the target is 5 orders-of-magnitude thermal noise rejection while providing a high signal efficiency at or above 95\%. 

Typically using a more complex network structure yields more accurate results, but this also creates a slower network. These two constraints need to be optimized as the deep learning architecture is developed. In the following two sections, deep learning techniques are used to train models then study their efficiency and processing time. In section \ref{section:cross_cor}, a commonly used method of template matching will be investigated to compare with the deep learning approach. 

\subsection{Generation of training data sets}
\label{sec:datasetgeneration}
NuRadioMC \cite{NuRadioMC} is used to simulate a representative set of the expected neutrino events for the ARIANNA detector, following the same setup as described in Sec.~\ref{subsection:sensitivity} but for randomly distributed neutrino energies that follow an energy spectrum expected for an astrophysical and cosmogenic neutrino flux; the astrophysical flux measurement by IceCube with a spectral index of $\gamma=2.19$ \cite{IceCubeICRC2017} is combined with a model for a GZK neutrino flux \cite{GZK_flux,GZK_2019} based on Auger data for a 10\% proton fraction \cite{van_Vliet_2019}. 

The resulting radio signals are simulated in the four LPDA antennas of the ARIANNA station by convolving the electric-field pulses with the antenna response, and the rest of the signal chain is approximated with an \SI{80}{MHz} to \SI{800}{MHz} band-pass filter. An event is recorded if the signal pulse crossed a high and a low threshold of 3.6 times $V_{\text{RMS}}^{\text{noise}}$ within \SI{5}{ns} in at least two LPDAs within \SI{30}{ns}. At such a low trigger threshold, noise fluctuations can fulfil the trigger condition at a non-negligible rate. Therefore, the signal amplitude is required to be at least 2.8 times the $V_{\text{RMS}}^{\text{noise}}$ before adding noise to avoid spurious triggers on thermal-noise fluctuations. In total 121,597 events that trigger the detector are generated and this is called the \emph{signal data set} in the following. 

The training data set for thermal noise fluctuations is obtained by simulating thermal noise in the four LPDA antennas and saving only those events where a thermal noise fluctuation fulfills the trigger condition described above. In total 1.1 million events are generated and this is called the \emph{noise data set} in the following.  

The limitations of the simulations and their impact on the obtained results are discussed at the end of this article.

\subsection{Network structures and training} 
\label{subsection:dl_eff_plot}
All of the networks are created with Keras \cite{chollet2015keras}, a high-level interface to the machine-learning library TensorFlow \cite{tensorflow2015-whitepaper}. Our primary motivation is to develop a thermal noise rejection method that operates on the existing ARIANNA hardware with an evaluation rate of at least \SI{50}{Hz}, which is a factor of $10^{4}$ larger than our current trigger rate. To increase the execution rate of the neural network, the hardware is one option to optimize; however, any alteration to the hardware is constrained by two main factors: the power consumption of the component and the reliability in the cold climate. Thus, this study will focus primarily on optimizing the execution rate by identifying the smallest network that reaches our objective. While the number of trainable parameters can give an indication of network size, the number of Floating Point Operations (FLOPs) is the chosen metric for network size in this paper. The number of FLOPs can be approximated by multiplying the amount of (+,-,*,/) operations performed by floating point numbers with the amount of nested loop iterations required to classify incoming data.

Besides making the network size smaller, another  way to improve the network speed is to reduce the input data size. Instead of feeding the signal traces from all four antennas into the network, one way to cut down on the size of input data is to use only the two antennas that caused the trigger. As each signal trace consists of 256 samples, the total input size to the network is 512 samples. In addition, a further reduced input data set is studied for various sizes by selecting the antenna with the highest signal amplitude and only using a window of values around the maximum absolute value. The window size was not fully optimized, but a good balance between input data size and efficiency is 100 samples around the maximum value. The reasoning for this is that the dominant neutrino signal does not span over the whole record length and typically only spans over less than 50 samples. 

\begin{figure}
{\centering
\includegraphics[width=0.49\textwidth]{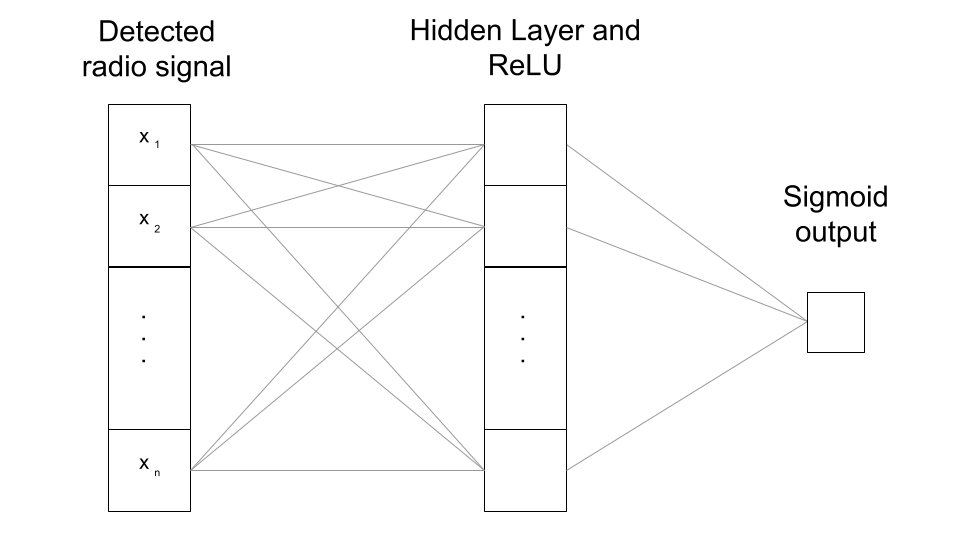}
\includegraphics[width=0.49\textwidth]{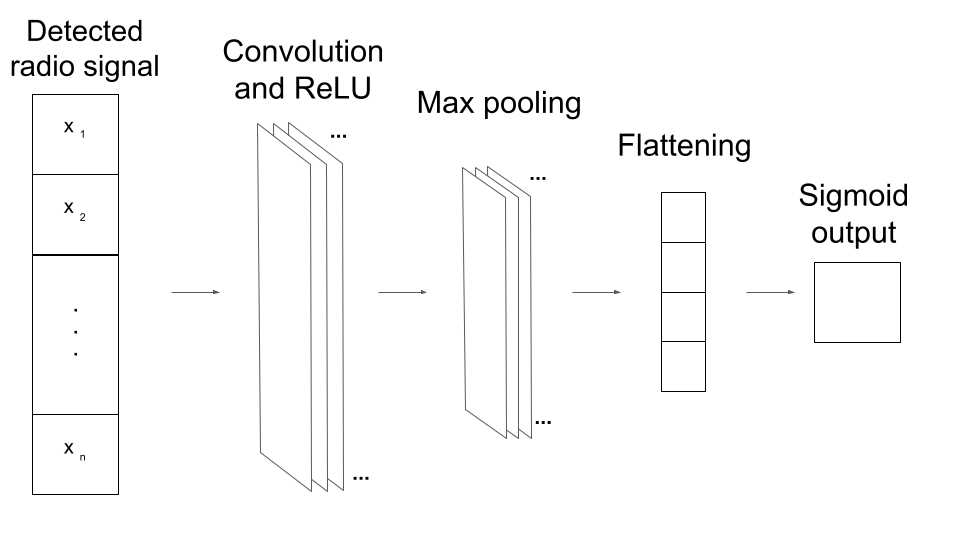}
\caption{Baseline architecture of a fully connected neural network (FCNN) on the left and a convolutional neural network (CNN) on the right. The FCNN contains one hidden layer with ReLU activation and a sigmoid activation in the output layer. The CNN in composed of a convolution with ReLU activation, max pooling, a flattening layer where the data are reshaped, and a sigmoid activation in the output layer.}
\label{fig:NN}
}
\end{figure}

The two network architectures studied in the following are a fully connected neural network (FCNN) \cite{Schmidhuber_2015} and a convolutional neural network (CNN) \cite{inproceedings,KIRANYAZ2021107398}, depicted in Fig.~\ref{fig:NN}.
The FCNN used in this baseline test is a fully connected single hidden layer network with a node size of 64 for the 100 input samples and 128 for the 512 input samples, a ReLU activation, and a sigmoid activation in the output layer. The CNN structure consists of 5 filters with 10x1 kernels each, a ReLU activation, a dropout of 0.5, a max pooling with size 10x1, a flattening step to reshape the data, and a sigmoid activation in the output layer. Both the CNN and FCNN are trained using the Adam optimizer with varying learning rates from 0.0005-0.001 depending on which value works best for each individual model. The training data set contains a total of 100,000 signal events and 600,000 noise events, where 80\% is for training and 20\% is to validate the model during training. Once the network is trained, the test data are used which contain 21,597 signal events and 500,000 noise events.

\subsection{Deep learning performance}
\begin{figure}
{\centering
\includegraphics[width=0.7\textwidth]{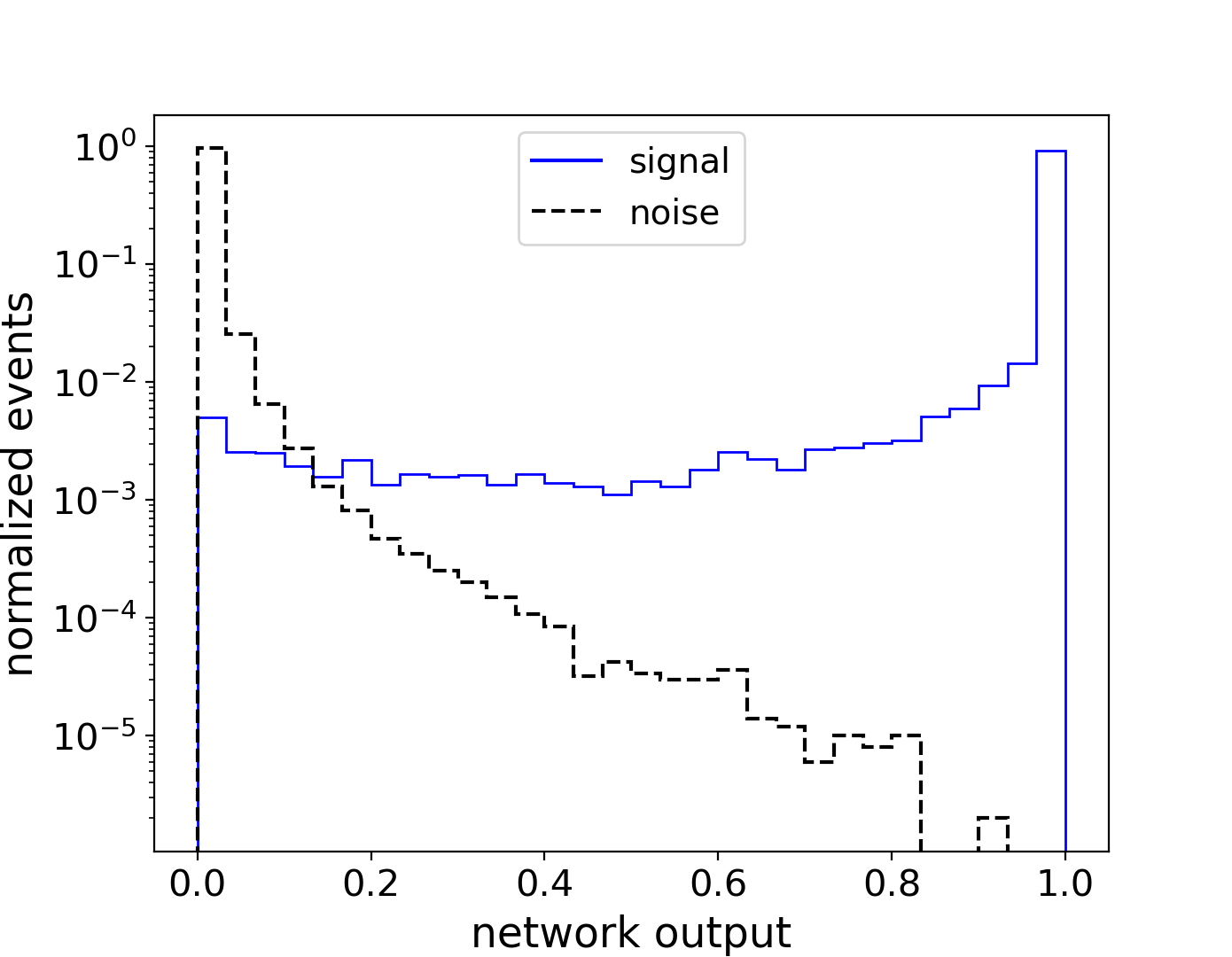}
\caption{Histogram of the network output for signal and noise classification. The network used for training and validation was a CNN with one convolutional layer comprised of 5 10x1 filters and input data of 100 samples around the maximum value of the waveform.}
\label{fig:prob_plot}
}
\end{figure}

\begin{figure}
{\centering
\includegraphics[width=0.7\textwidth]{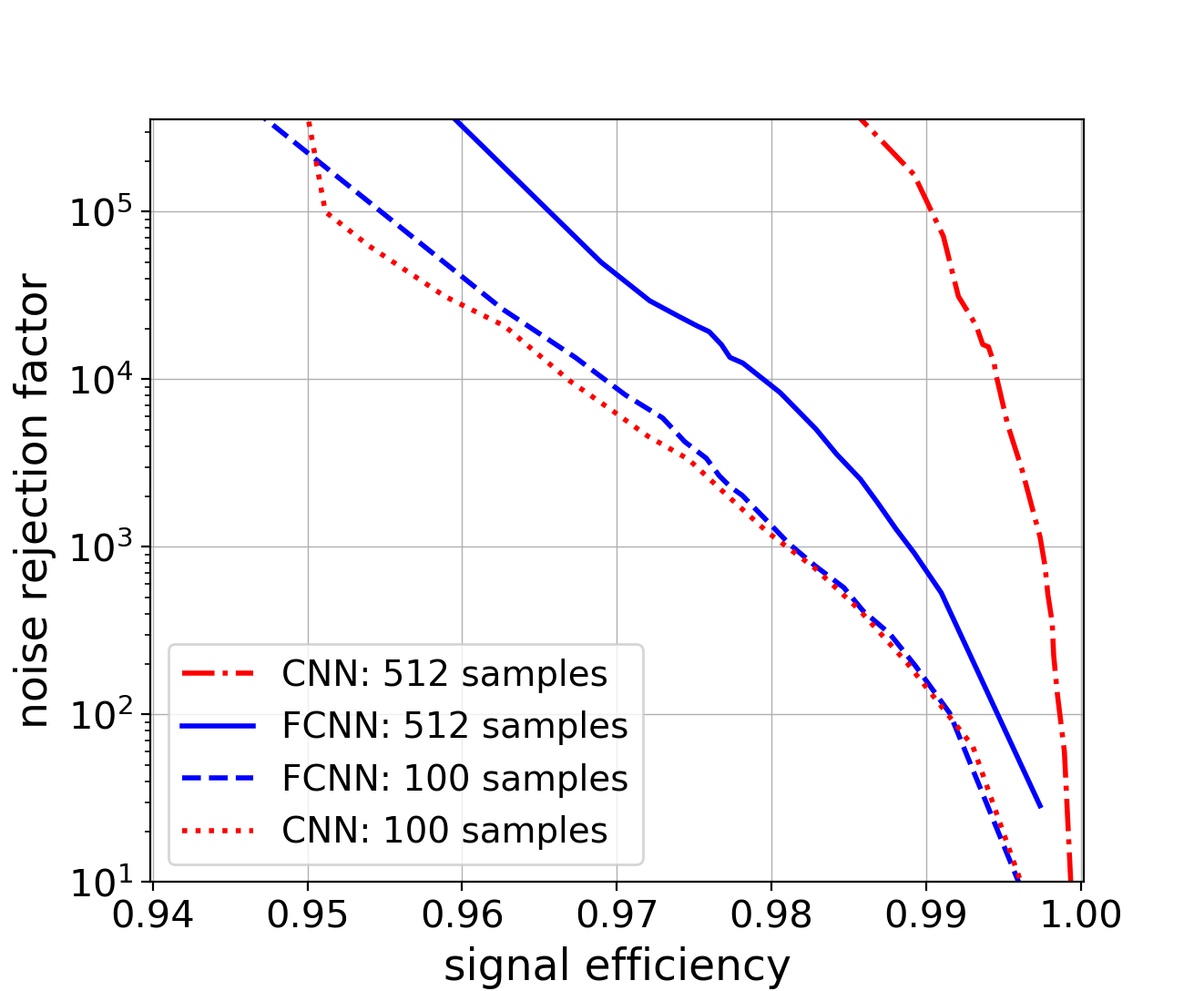}
\caption{Signal efficiency vs. noise rejection factor for FCNN's and CNN's with two different input data sizes (100 samples and 512 samples). Both CNN's have the structure of one convolutional layer containing 5 10x1 filters. The FCNN's have one fully connect layer with node size 64 for the 100 samples input data and node size 128 for the 512 samples input data. }
\label{fig:eff_plots}
}
\end{figure}

The signal and noise event classification score distributions from the networks are distinct. With the sigmoid activation in the output layer, the classification distribution falls between 0 and 1, where close to 0 is noise-like data and close to 1 is signal-like data. Once trained, with the 100 input sample CNN mentioned above, the distribution shown in Fig.~\ref{fig:prob_plot} is obtained. From this distribution, the amount of signal efficiency vs. noise rejection can be varied by choosing different network output cut values. Training and testing these networks with each input data size yields the signal efficiency vs. noise rejection plot in Fig.~\ref{fig:eff_plots}. Each data point corresponds to a different network output value, and the final cut value is chosen by optimizing the noise rejection for the desired signal acceptance. All of these input data sizes produce efficiencies above the required threshold of 95\% for signal, and all were able to reach at least 5 orders-of-magnitude noise rejection.

Since all of the networks have efficiencies above our target of 95\% for signal at $10^{5}$ noise rejection, the main consideration is the amount of FLOPs required for each network because this directly impacts the processing time. Typically, CNN's have less parameters overall due to their convolutional nature, which focuses on smaller features within a waveform; comparatively, the FCNN considers the whole waveform to make its prediction, so it requires more node connections. The next step is to investigate the FLOPs for each network, and determine the processing time on a given device.

\section{Processing time and reliability on devices}
\label{section:proc_time}
In this section, the processing time of the deep learning filter is studied. As the filter is intended as a real-time trigger, a fast execution time is crucial. The current ARIANNA hardware is used to test and measure the execution time under realistic conditions. 

There are several time components that impact the physics capabilities of the ARIANNA detector: (1) the time to transfer the data from the waveform digitizers to the microcomputer, $T_{\text{read}}$, (2) the time to reformat and calibrate the raw data for the deep learning evaluation, $T_{f}$, and (3) the time to evaluate the event with deep learning and make a decision, $T_{dl}$, (4) the time to store an event to the local SD card, and (5) the time to transmit the event via the Iridium satellite network. 
The architecture of the ARIANNA pilot station cannot acquire neutrino events while the data acquisition system is processing events or during the transmission of data over Iridium satellite. It is useful to express the processing time in terms of fractional loss of operational time, or deadtime. Operational livetime, L, is the calendar time of nominal operation, T, corrected for the time losses due to event processing and transmission. Often, the operational livetime is reported as a fractional quantity, $f_L=L/T$ and the fractional operational deadtime, $f_D$=1-$f_L$.  

If the rate of saving events to the SD card is sufficiently small, then the time to process an event using deep learning is given by $T_{\text{min}}$=$T_{\text{read}}$+$T_{f}$+$T_{dl}$. As shown in this section, the latter two time scales depend on the microprocessor. The time to transfer data to the microprocessor depends on the details of the data acquisition system. This is known for the pilot ARIANNA stations, $T_{\text{read}}=$ \SI{7.3}{ms}, but the design of this station did not focus on minimizing $T_{\text{read}}$, and this value can be reduced by redesigning the hardware. It is assumed to be negligible when evaluating new platforms for future designs of the ARIANNA data acquisition system.

\subsection{Processing time}
Two microprocessors are explored for their processing time and power consumption: a Raspberry Pi compute module 3+ microcomputer and an MBED LPC1768 ARM microcontroller. The MBED is the current device installed in ARIANNA and is implemented through custom C code. The Raspberry Pi is a microcomputer with a Raspbian operating system, which is based on Debian. As with the MBED, the neural network is implemented with a similar custom code on the Raspberry Pi. Since the optimal networks found in the previous section are small and shallow, a custom code is written that implements the trained neural networks in C for maximum performance. To test the prediction capabilities and the classification time in both devices, a simulated event is read in and either matrix multiplied by the array of weights and biases in the FCNN case or convolved with the weights and bias filters in the CNN case. 

Two methods are used to measure the MBED and Raspberry Pi processing times, $T_{dl}$. For the Raspberry Pi, since it is not attached to the ARIANNA data acquisition system (DAQ), the processing time is measured by looping over the processing code 100 times, while measuring the total time for 100 loops with the clock function in C. The total time divided by 100 is the average processing time per event. For the MBED, since it is attached to the ARIANNA board, it has the ability to be probed for reset pulses. Reset pulses are used by the MBED to reset the logic of the FPGA and triggering circuitry to prepare for a new event. The time between reset pulses will provide the total deadtime. In the case that the station is triggered continuously, which would result in 0\% livetime, the time between reset pulses corresponds to the processing time. Livetime is defined as the time between the reset pulse of the previous event and the trigger of a new event. To accomplish this setup, a pulse is injected into the hardware with an amplitude large enough to trigger the system. By increasing the injection rate until the system experiences 0\% livetime, the minimum time difference between reset pulses, $T_{\text{min}}$, can be found. 

$T_{\text{min}}$ can be broken down into three time components that when summed together make up the processing time, $T_{\text{min}}$=$T_{\text{read}}$+$T_{f}$+$T_{dl}$. To measure each component, different processing functions are disabled. For example, when the deep learning code and reformatting code are disabled, $T_{\text{min}}$=$T_{\text{read}}$. The processing time $T_{dl}$ and reformatting time $T_{f}$ for both devices are shown in Table~\ref{table:MBED_vals} for a given model along with the FLOPs of each network. While Table \ref{table:MBED_vals} shows that the relationship is not completely linear, FLOPs provide a reasonable proxy to estimate the relative speeds of specific deep learning models. All models listed reach the required efficiency of 95\% neutrino signal at $10^5$ noise rejection. Therefore, the fastest network is chosen for the lab tests, which is the CNN with 100 input samples.

\begin{table}[h!]
\centering
\begin{tabular}{ c c c c c } 
 \hline  \hline
 Variable& model & FLOPs & MBED  &  Raspberry Pi 
 \\ 
 \hline
 &FCNN 512 samples &131,457& \SI{45}{ms}  & \SI{2.5}{ms}
 \\ 
 &CNN 512 samples &55,816& *  & \SI{1.5}{ms}
 \\
  $T_{dl}$&FCNN 256 samples &32,961& \SI{13}{ms} & \SI{1.0}{ms} 
 \\ 
 &CNN 256 samples &27,376& \SI{9.4}{ms} & \SI{0.95}{ms}
  \\
 &FCNN 100 samples &12,993& \SI{4.7}{ms} & \SI{0.46}{ms}
 \\ 
 &CNN 100 samples &10,096& \SI{3.7}{ms} & \SI{0.39}{ms} 
 \\ 
 \hline
 $T_{f}$&all networks & & \SI{1.3}-\SI{1.9}{ms} & \SI{0.095}-\SI{0.12}{ms} 
 \\ 
 \hline \hline
\end{tabular}
\caption{Processing times per event, $T_{dl}$ and the number of Floating Point Operations (FLOPs) of various models that demonstrate the required efficiency, and the reformatting time per event $T_{f}$ for 100 and 512 input data sizes respectively, for an MBED and a Raspberry Pi. *memory limitations prevented this measurement.}
\label{table:MBED_vals}
\end{table}

\subsection{Reliability and power consumption}
Optimizing the network architecture and processing time are not the only factors to consider when implementing a deep learning network onto ARIANNA. Reliability in the harsh Antarctic climate must be considered as well as the limited power available in the remote location of Antarctica. The MBED was tested in the field for reliability in cold temperatures (averaging \SI{-60}{\degreeCelsius}) and meets the specification on power consumption, operating under one Watt. In contrast, the Raspberry Pi is rated to \SI{-25}{\degreeCelsius} \cite{raspberrypi_temp} and requires more power than the MBED. The Raspberry Pi was stress tested under cold conditions, running it from \SI{20}{\degreeCelsius} to \SI{-60}{\degreeCelsius}. It ran continuously with the deep learning filter for the hour it took to cool down to \SI{-60}{\degreeCelsius}, and then it was run for an additional hour at this temperature. Once it was brought back up to room temperature, the Raspberry Pi was still operational. If chosen as the optimal device in the future, the Raspberry Pi would need to be tested further for long term operational reliability such as temperature cycling. 

Due to the extensive work required to implement the deep learning network on a new device, for this paper the first analysis is done with the current MBED microcontroller. As discussed in the previous section, the 100 input sample CNN is chosen for the experimental verification study. Next, the template matching method is applied to the same simulated data set to compare with the 100 input sample CNN. Then the 100 input sample CNN is implemented into the current MBED software for experimental verification.

\section{Performance verification}

In this section, the performance of a deep learning filter is compared to the commonly used template matching, verified in a lab measurement, the ARIANNA hardware's computing performance is studied, and measured in situ cosmic-ray data are classified by the filter to study signal performance. 

\subsection{Comparison to template matching}
\label{section:cross_cor}
We compare the performance of a deep learning filter to a realistic template matching procedure using a single template, similar to what was used in a previous analysis \cite{Anker:ARIANNAlimit2019}. It is found that the deep learning method is typically faster and performs better. 

\begin{figure}
    \includegraphics[width=\textwidth]{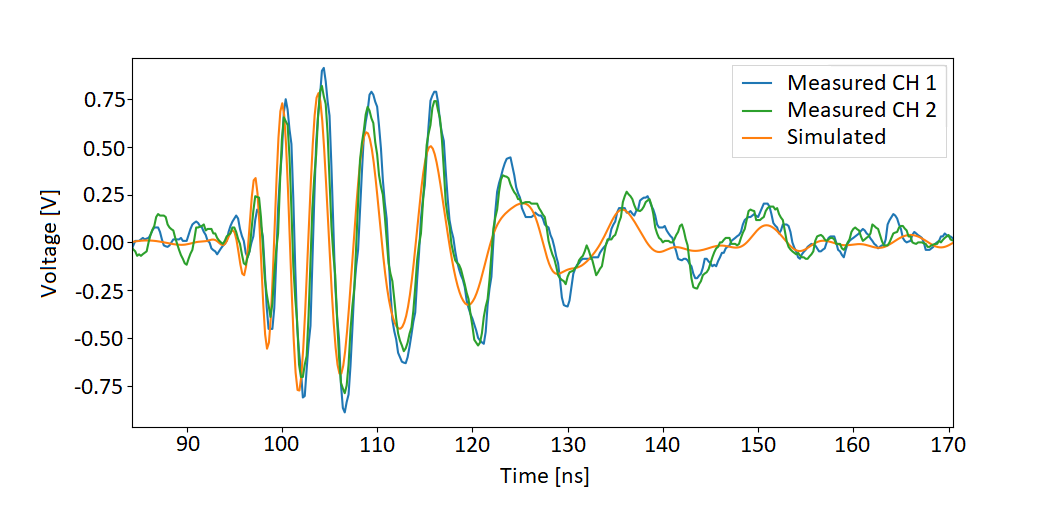}   
    \caption{Comparison of the analog/measured neutrino template signal being produced by the pulse generator and the simulated neutrino template. For details on the pulses see Sec.~\ref{subsection:exp_verif}.}
    \label{fig:pulse_comparison}
\end{figure}

\begin{figure}
{\centering
\includegraphics[width=0.65\textwidth]{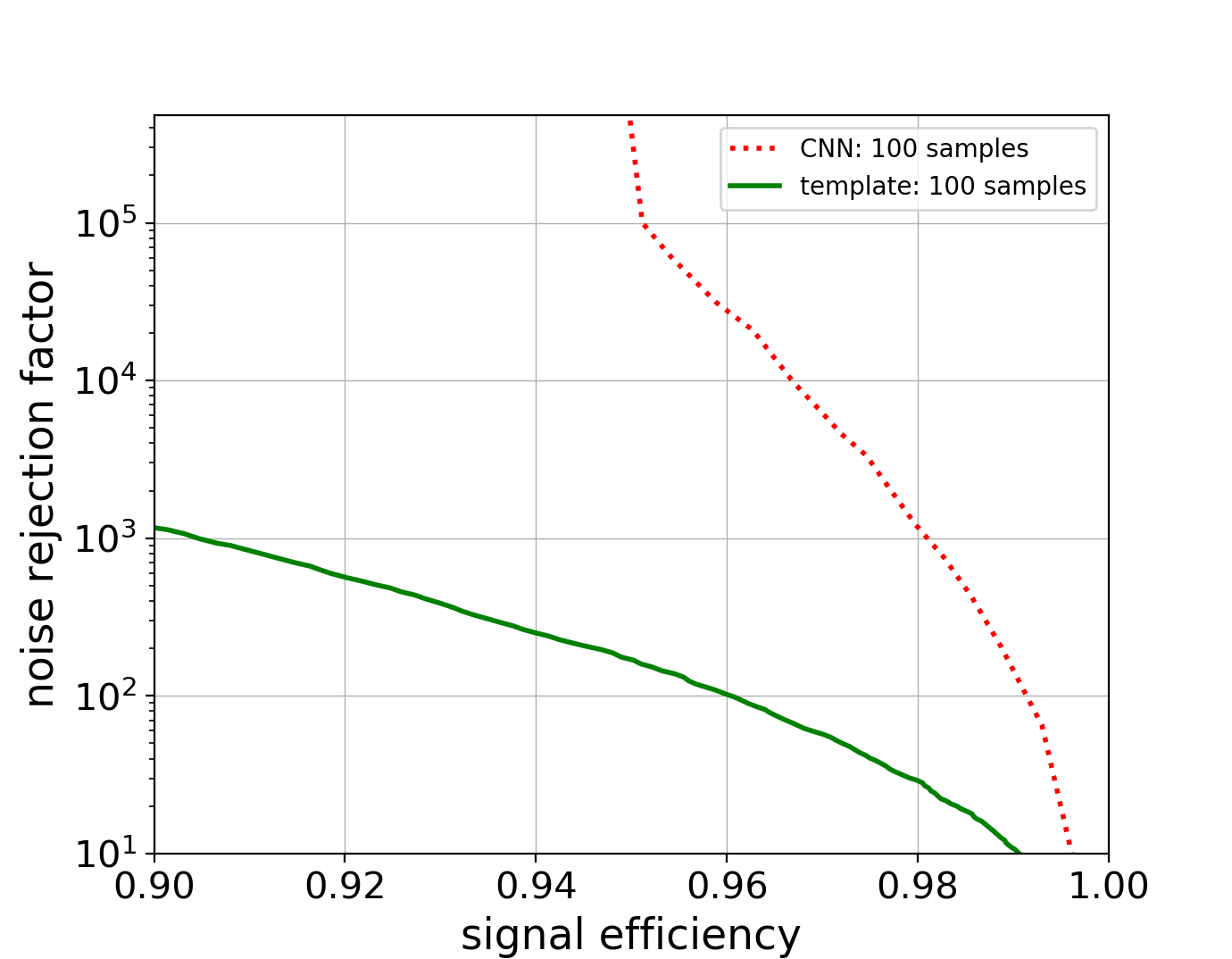}
\caption{Signal efficiency vs. noise rejection factor for the 100 input sample CNN and the 100 samples template matching method. The template matching technique uses a simulated neutrino template (with antenna and amplifier response, and no noise) to perform a cross-correlation on the same simulated data used to train the 100 input sample CNN.}
\label{fig:templ_eff_plot}
}
\end{figure}

A neutrino template is constructed by convolving a predicted Askaryan pulse with the antenna, amplifier and filter responses of the ARIANNA signal chain as already done in previous analyses \cite{Anker:ARIANNAlimit2019}. A single template is used to minimize the computational costs, and also because of the observation that the template is dominated by the detector response; variations in the predicted Askaryan pulse have only a small influence on the resulting templates (see e.g. \cite{Anker:ARIANNAlimit2019, RNOGEnergy2021}). The plot of the general simulated template waveform is found in Fig.~\ref{fig:pulse_comparison}, and for this study, the amplifier response is added to this waveform without noise. Following the same data format as the 100 input sample CNN, the template was trimmed to 100 samples around the maximum absolute value of the waveform. This template was cross-correlated with the simulated signal and noise data sets, and the maximum absolute value of the cross-correlation is used as a measure for signalness, i.e., the output is a number between 0 (noise-like) and 1 (signal-like) as in the deep learning case. To compare the performance of the template and neural network method, the signal efficiency vs. noise rejection factor is computed and compared to the CNN result which is presented in Fig. ~\ref{fig:templ_eff_plot}. 

The CNN method is found to perform significantly better. At the benchmark value of 95\% signal efficiency, the template method only achieves a little more than two orders-of-magnitude noise rejection. One explanation for this is that the CNN identifies smaller 10x1 features within the training sets, which gives it more flexibility. Additionally, the CNN has 5 times the amount of features to extract compared to the template's single waveform/feature. Another aspect of the template matching technique is to determine the processing speed. Estimating the processing speed for this method, the FLOPs are roughly 29,900, which is close to three times the amount of FLOPs of the 100 input sample CNN. Narrowing in further on the template signal pulse to 50 samples around the maximum of the waveform, the FLOPs are roughly 7,450. This is now less FLOPs compared to the 10,096 FLOPs of the 100 input sample CNN, but the efficiency of the template matching is still significantly worse. Therefore, the cross-correlation neutrino template matching method is less efficient and (depending on the input data size) slower than the CNN technique. 

\subsection{Laboratory verification}
\label{subsection:exp_verif}
The deep learning filter uses the 100 input sample CNN described in section \ref{sec:dl_arch}. Data taking with LPDA antennas proves a challenge in the lab due to the lingering radio frequency noise present in the environment. Without a radio quiet space, one cannot replicate the environment of the Antarctic ice since the antennas would measure local radio sources which would bias the data. Thus, for in-lab tests of the deep learning implementation, an experimental "post LPDA antenna" radio neutrino pulse is created and injected into the ARIANNA hardware, bypassing the antenna, to verify the simulated results.

\begin{figure}
    \centering
    \includegraphics[width=1.0\textwidth]{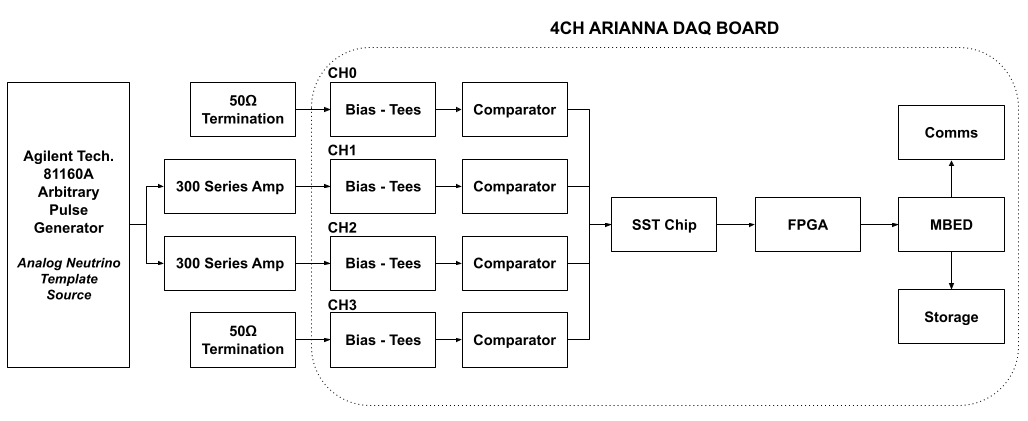}
    \caption{Diagram showing the set up for the collection of measured signal. The neutrino template was loaded onto the Agilent Tech. Arbitrary Pulse Generator and converted to an analog signal. The analog neutrino template was then injected into a series 300 amplifier then into the ARIANNA DAQ board which contains triggering circuitry, an SST Chip, FPGA, and MBED.}
    \label{fig:DAQDia}
\end{figure}

The expected neutrino template is generated with the standard simulation tool, NuRadioMC \cite{NuRadioMC}, which convolves the expected electric field at the detector with the LPDA antenna response. This neutrino template is then programmed into an Agilent Tech. 81160A arbitrary pulse generator to produce an analog version of the neutrino template as observed by the LPDA antenna. This waveform is then injected into the Series 300 amplifier of the ARIANNA pilot station, which adds realistic amplifier noise. By adjusting the input amplitude of the template, the SNR can be tuned to arbitrary values. Figure~\ref{fig:DAQDia} shows a diagram of the experimental set up. The noisy signal is then routed to the input of the ARIANNA DAQ for data taking. Figure~\ref{fig:pulse_comparison} gives a comparison of the simulated neutrino template to those produced by the analog pulse generator, known  as measured neutrino template.

The DAQ consists of an SST chip, an FPGA, and an MBED microcontroller. Once digitized by the SST chip, the FPGA passes the digitized data (event) to the MBED, where the channel with the largest signal is chosen and runs through the deep learning filter. Once finished, the filter gives the network output value of the event to determine its classification. Any event whose probability is below the cutoff would normally not be saved to memory. For testing purposes, all events are saved into memory along with the deep learning calculated probability. The deep learning filter performance on the ARIANNA hardware is also checked on a local computer which takes the digitized data and recalculates the probability. Both methods show equivalent results.  

\begin{figure}
    \centering
    \includegraphics[width=0.7\textwidth]{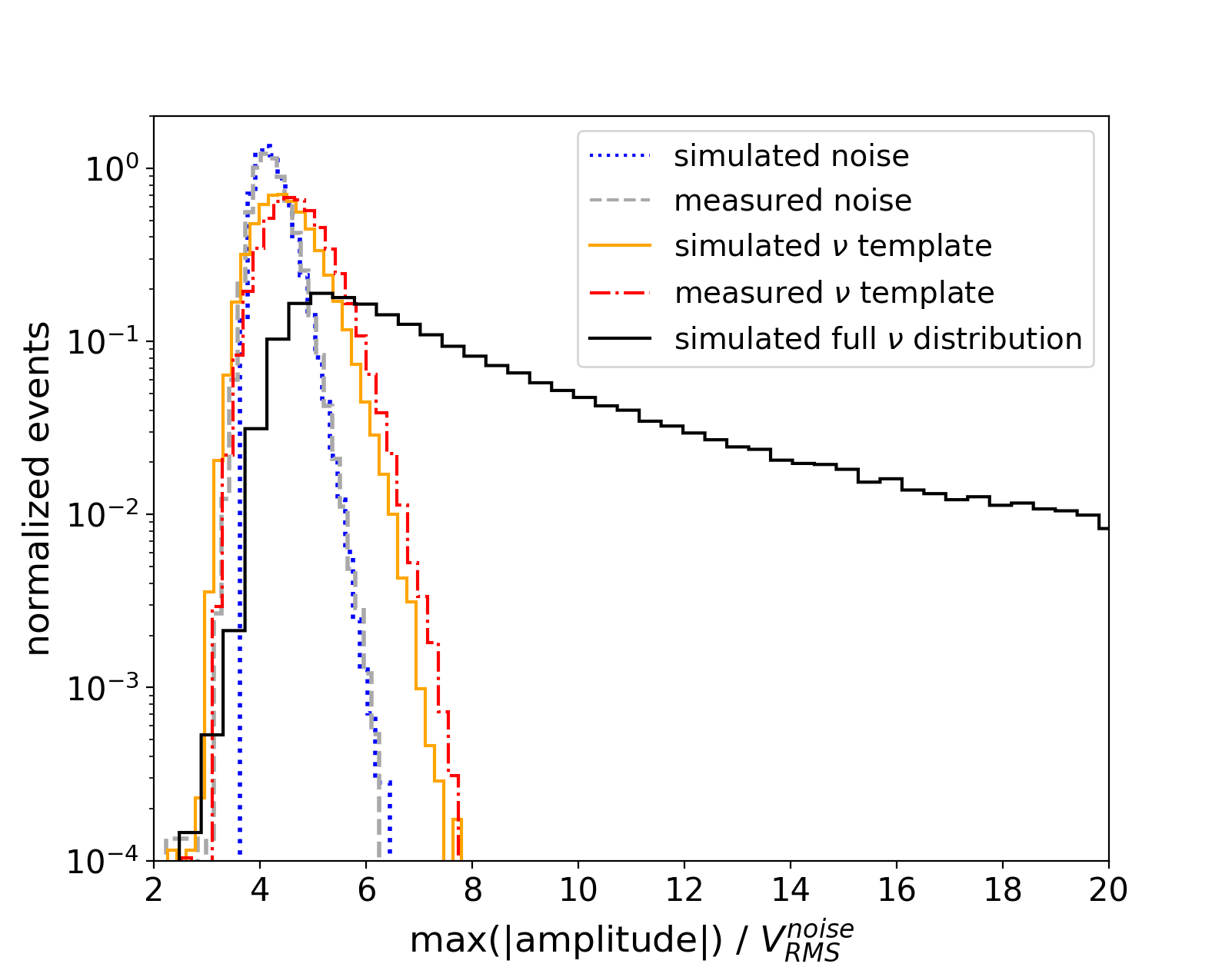}
    \caption{Histogram of the Signal to Noise Ratio (defined here as the ratio of the maximum absolute value of the maximum waveform to the noise RMS, $V_{\text{RMS}}^{\text{noise}}$) distributions for simulated thermal noise, measured thermal noise, simulated singular neutrino template, measured singular neutrino template, and simulated full neutrino spectrum. The template data have Gaussian noise added to the template to get the distribution above. The full neutrino distribution data set was used to train all of the networks in section \ref{sec:dl_arch}. }
    \label{fig:SNRs}
\end{figure}

This setup allows for the recording of neutrino signal data sets and noise data sets. To record the noise data set, the neutrino signal generator is deactivated so that the ARIANNA DAQ only sees thermal noise amplified by the amplifiers. This test focuses on low SNR events which are the most difficult to differentiate between signal and noise. Therefore, the high/low trigger threshold is set to an SNR of 3.6 and a two of four coincidence logic between channels is required. The corresponding SNR distribution is shown in Fig.~\ref{fig:SNRs}, where the SNR is calculated as the maximum amplitude of any of the channels divided by the RMS noise. The measured noise distribution matches well to the SNR distribution of the simulated noise data set used in the previous sections. 

Recording the neutrino signal data set is more challenging because at these low thresholds, a noise fluctuation might trigger the readout. So instead of a self-trigger, the ARIANNA DAQ is externally triggered via the trigger output of the pulse generator. The time delay is adjusted so that the signal pulse is at the correct location as expected from the ARIANNA high/low trigger. The amplitude of the signal pulse is adjusted to produce a low SNR distribution just above the trigger threshold. The neutrino template in Fig.~\ref{fig:pulse_comparison} remains constant for all of the tests, but since the noise added by the amplifier follows a Gaussian distribution, the resulting SNR follows the same distribution. As seen in Fig.~\ref{fig:SNRs}, the distribution is broader and shifted to a slightly higher mean than the noise data sets. This is expected because the interference of the extended signal pulse with noise gives several chances for an upward fluctuation, and the maximum value of any channel is chosen for the SNR estimate. The resulting SNR distribution of this low-amplitude simulated neutrino signal data set matches the experimental distribution.

\begin{figure}
    \centering
    \includegraphics[width=0.7
    \textwidth]{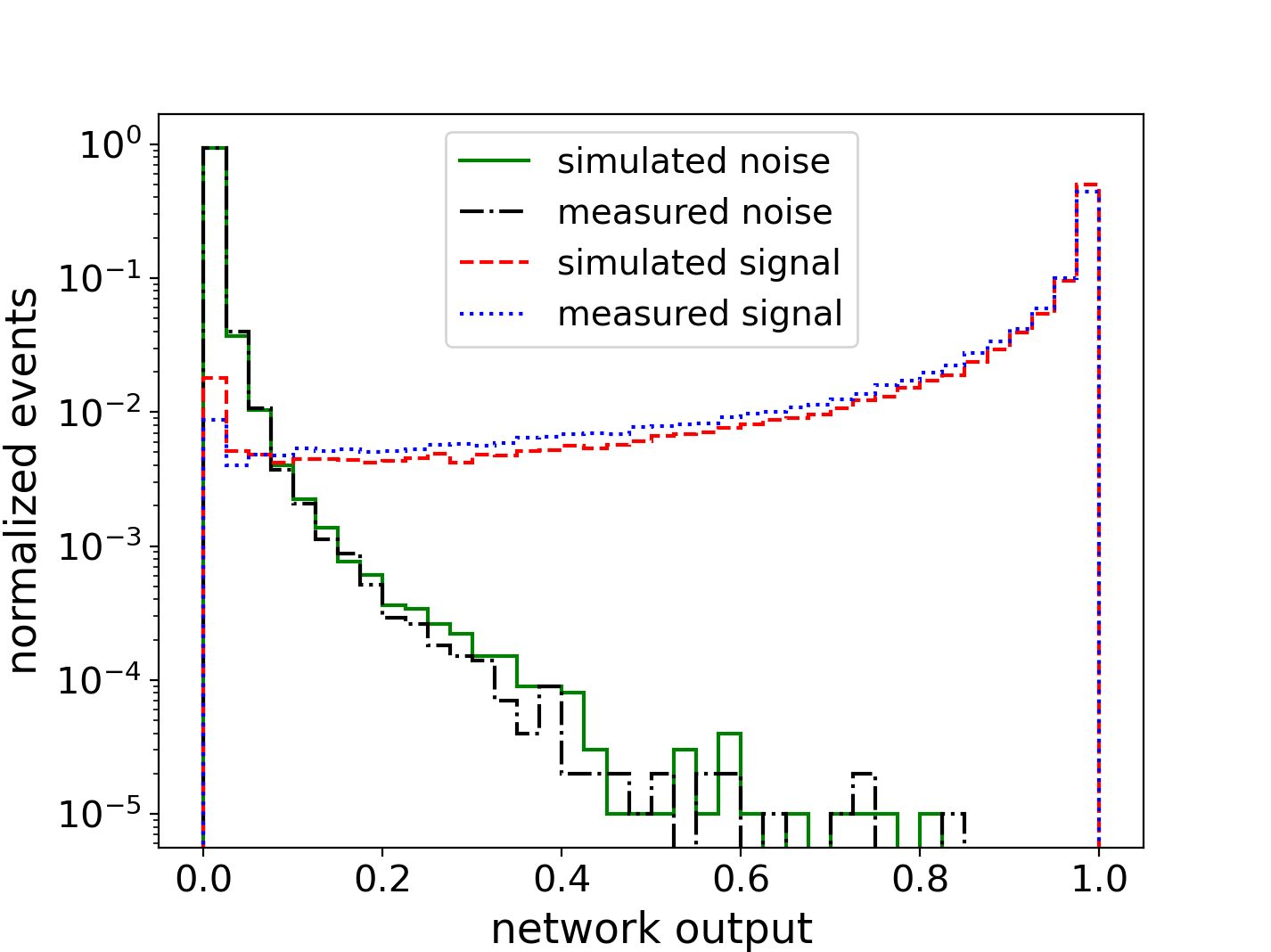}
    \caption{Histogram of the network output values of simulated and measured signal template and noise. The values are obtained from the network's output layer (a sigmoid activation function).}
    \label{fig:probs}
\end{figure}

To test the deep learning filter and verify that the simulated hardware components are comparable between measured and simulated data, a histogram of the network output values is plotted. This distribution is obtained by using the trained 100 input sample CNN to classify simulated and measured events. As seen in Fig.~\ref{fig:probs}, the distribution of the simulated MC data set agrees well with the experimentally measured distributions. The small deviations are are likely due to differences in SNR distributions and environmental effects such as strong radio pulses leaking into the cables. This gives us confidence that the MC simulation indeed describe measured data and that the inferred noise rejection factor and signal efficiency are credible. 

\subsection{ARIANNA hardware computing speed for the CNN filter}
\label{subsection:hardware_performance}
The deep learning filter will increase the low-level trigger rate to \SI{100}{Hz}. This significant change in trigger rate raises the question of whether or not the current hardware will be able to handle triggering effectively at \SI{100}{Hz}. This section explores the hardware computing performance of the ARIANNA pilot station, which includes the waveform digitizers, the readout into the FPGA, the transmission from FPGA to the MBED for further processing, and the application of the deep learning filter. Taking the values from Table~\ref{table:MBED_vals}, and now including $T_{\text{read}}=$ \SI{7.3}{ms} (which was experimentally obtained with the same procedure as $T_{dl}$), the MBED has the event readout time $T_{\text{min}}=$ \SI{12.3}{ms}. 

This deadtime is the total processing time of transferring and packaging the data, and running on-board analysis programs. Deadtime is acquired by measuring the time delay between reset pulses, which is sent by the MBED when it is ready to receive a new event. To study the effect of $T_{\text{min}}$ on operational livetime, the ARIANNA triggering system is simulated. $T_{min}$ is directly related to operational livetime. First the effect of the deadtime on the triggering system is simulated by writing code that takes an exponential distribution of times, dt, representing the time between incoming triggers  with a mean of $R_{T}^{-1}$, where $R_{T}$ is the noise trigger rate. When an event is being processed no other triggers can be accepted until after the previous event is finished processing. The simulation uses the distribution of times to calculate a fractional livetime as a function of $R_{T}$ by calculating the average portion of time spent in deadtime due to $R_{T}$ over a time T.

\begin{figure}
    \centering
    \includegraphics[width=0.75
    \textwidth]{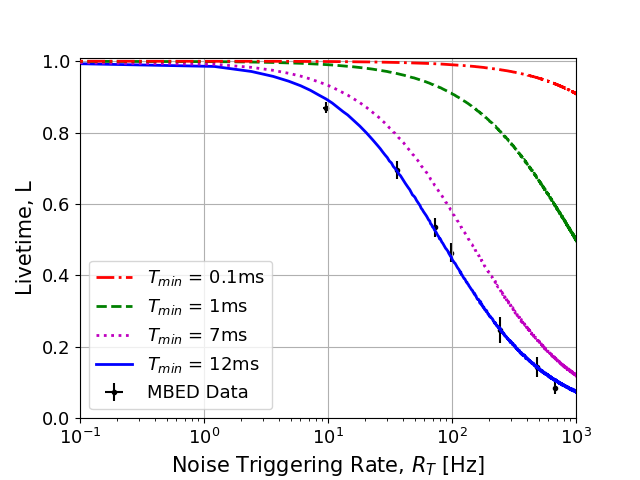}
    \caption{Livetime, L, as a function of Noise Trigger Rate, $R_T$, for three assumptions on the instrumental deadtime. Also plotted are data (black dots) from experimental verification study.}
    \label{fig:LTvsNoiseTrigg}
\end{figure}

To confirm the simulation results for the fractional livetime as a function of noise trigger rate, the ARIANNA DAQ board is injected with signal-like voltage pulses periodically at a rate of \SI{100}{mHz} and a high enough amplitude that always fulfills the trigger condition. At the same time, the trigger threshold is lowered after every trial to increase the noise trigger rate. For example, at a noise triggering rate of \SI{0}{Hz}, 100\% fractional livetime is expected since the injection rate of the periodic pulses is greater than the deadtime, allowing all injected pulses to be processed and saved to the SD card. In Fig.~\ref{fig:LTvsNoiseTrigg}, the results of the experiment can be seen as the threshold is lowered, which in turn increases the noise rate and decreases the livetime. The measurement is described well by the corresponding simulation for $T_{\text{min}} = \SI{12}{ms}$.

At \SI{100}{Hz} noise triggering rate, the livetime decreases to an unacceptably low value of below 50\%. The CNN reaches a rejection factor of $10^{5}$, but trigger rates are limited to \SI{10}{Hz} or lower to keep instrumental deadtime under 10\%. This result motivates the consideration for improvements in hardware, such as replacing the MBED with a faster processing device.

\subsection{Classification of cosmic-ray data}
\label{subsection:CR}
The ARIANNA detector stations are simultaneously sensitive to cosmic rays that interact in the atmosphere and neutrinos; in addition to the downward facing antennas that are designed to detect neutrinos, the ARIANNA detector contains upward facing antennas to detect cosmic rays. Cosmic rays provide a calibration beam with similar radio frequency content and time variation as expected from neutrinos \cite{Barwick2017-Airshowers}. Moreover, both cosmic rays and neutrinos have bipolar signals that are short compared to the response time of the antennas, and their signals are significantly distinct from thermal fluctuations. Therefore, they provide another opportunity to verify the behavior of the CNN on signals that should not be rejected as noise. Cosmic rays trigger a given station at a rate of only a few events per day, so it is important to keep as many cosmic-ray events as possible.

To test the network, a set of cosmic-ray events from ARIANNA station 52, collected between November 2018 and March 2019, are used \cite{leshanICRC}. For this data, the two of four coincidence logic trigger is set to a threshold of 4.4 SNR, and the event set has an expected purity of >95\%. Then the 100 input sample CNN trained on neutrino data and described in section \ref{sec:dl_arch} is used on these cosmic-ray data. This network classifies 102 out of 104 cosmic-ray events as signal, which is a greater value than neutrino efficiency in this study because the thresholds are slightly larger (4.4 SNR compared to 3.6 SNR in the simulation study). 
A network trained only with neutrinos still identifies cosmic rays adequately. It is likely that an even better cosmic-ray efficiency can be obtained by including them in the training data for a more robust CNN filter.

\section{Summary, discussion, and future plans}
Due to the low neutrino flux at ultra-high energies, the sensitivity of the detector to a flux of high-energy neutrinos is limited by statistics. Probing new parameter spaces is made possible by implementing deep learning techniques to increase the sensitivity of the ARIANNA detector. A small convolutional neural network (CNN) was implemented on the ARIANNA MBED microcontroller to discriminate between thermal noise fluctuations and neutrino signal. It was shown that CNN filters were much more accurate and computationally faster than simple cross-correlation methods (template matching) in distinguishing between thermal events and neutrino signals. Only one thermal event in every $10^5$ thermal triggers was misidentified by the CNN, while 95\% of the neutrino signal was correctly identified. Consequently, the trigger rate can be increased by five orders-of-magnitude while transmitting at an event rate of \SI{0.3}{mHz} over the Iridium communication network. This results in an increase in neutrino sensitivity of 40\% at \SI{e18}{eV} and up to a factor of two at lower energies. The simulation study was verified by lab measurements that found an excellent agreement between the measured and simulated distributions of both neutrino and thermal noise events. Then, a group of measured cosmic rays were run through the 100 input sample CNN, and 102 out of 104 events were classified as signal.

At the moment, the processing speed of the ARIANNA hardware is the limiting factor and restricts the low-level trigger rate to smaller than \SI{10}{Hz} to avoid deadtime. In the future, several improvements to the ARIANNA hardware will be considered. First, the ARIANNA hardware that sends the digitized data from the FPGA to the microprocessor can be parallelized to decrease the time to readout an event, which would provide the opportunity to trigger the detector at even higher rates. Second, more capable computing on the ARIANNA station through improved electronics will be studied. The current generation of ARIANNA hardware is now more than 10 years old, and many recent microcomputer systems offer more performance at comparable power consumption. An increase in computer speed allows more complex CNN architectures with an equivalent improvement in the trade off between neutrino signal efficiency and background rejection. The combination of these two changes would increase the sensitivity of the ARIANNA detector even if the communication transfer rates remained the same. However, the next generation of Iridium satellites, Iridium Next \cite{IridNext} has been recently deployed. This system has the potential to increase the transfer rates by many orders-of-magnitude relative to the SBD message transfer system currently used by ARIANNA.     

Reliance on deep learning filters may lead to unwanted results when incoming data deviates from training data, so they must be carefully evaluated by laboratory and field studies. The lab measurements described in this paper are encouraging, and suggestive that the simulations describe reality. The next stage of confirmation studies follow a similar plan that was used to validate the simulation studies of the sensitivity of the ARIANNA detector. After modifying the software in the data acquisition system to include the CNN filter, we plan to (1) use a variety of radio transmitters within a preexisting borehole drilled to a depth of \SI{1700}{m} at the South Pole (the SPice core \cite{SPICEcore}), to confirm signal efficiency at various thermal noise trigger rates, and (2) compare the rate and physical properties of cosmic-ray events to data samples collected without the filter. 

In addition to the current hardware constraints, there are limitations on the simulation tools as well. This analysis was able to use simulated data to achieve 5 orders-of-magnitude thermal noise rejection, but the simulations do not yet include a number of real world effects that can create small corrections to the data. One example is in the ice model, which does not include the layering structure of the ice density (stratification) when signals propagate through it. The incompleteness of the simulations can affect the in-field performance. However, these are known limitations that will be explored in the next round of simulations.

\section{Acknowledgement}
We are grateful to the U.S. National Science Foundation-Office of Polar Programs, the U.S. National Science Foundation-Physics Division (grant NSF-1607719) for supporting the ARIANNA array at Moore's Bay, and NSF grant NRT 1633631. Without the invaluable contributions of the people at McMurdo Station, the ARIANNA stations would have never been built. We acknowledge funding from the German research foundation (DFG) under grants GL 914/1-1 and NE 2031/2-1, the Taiwan Ministry of Science and Technology, the Swedish Government strategic program Stand Up for Energy, MEPhI Academic Excellence Project (Contract No.  02.a03.21.0005) and the Megagrant 2013 program of Russia, via agreement 14.12.31.0006 from 24.06.2013. The computations and data handling were supported by resources provided by the Swedish National Infrastructure for
Computing (SNIC) at UPPMAX partially funded by the Swedish
Research Council through grant agreement no. 2018-05973.

\bibliographystyle{JHEP}
\bibliography{bib}
\end{document}